\documentclass[conference]{IEEEtran}
\IEEEoverridecommandlockouts
\usepackage{cite}
\usepackage{amsmath,amssymb,amsfonts}
\usepackage{algorithmic}
\usepackage{graphicx}
\usepackage{textcomp}
\usepackage{xcolor}
\usepackage{authblk}
\usepackage{hyperref}
\usepackage{float}

\def\BibTeX{{\rm B\kern-.05em{\sc i\kern-.025em b}\kern-.08em
    T\kern-.1667em\lower.7ex\hbox{E}\kern-.125emX}}
\begin{document}

\title{Enhancing Emergency Communication for Future Smart Cities with Random Forest Model\\
}

\author[ ]{%
  \begin{minipage}[t]{0.45\textwidth}
    \centering
    Chengkun Ye \\
    School of Computer Science \\
    University of Nottingham, England \\
    \textit{alycy14@Nottingham.ac.uk}
  \end{minipage}%
  \hspace{0.1\textwidth} 
  \begin{minipage}[t]{0.45\textwidth}
    \centering
    Milena Radenkovic \\
    School of Computer Science \\
    University of Nottingham, England \\
    \textit{milena.radenkovic@nottingham.ac.uk}
  \end{minipage}%
}
\maketitle

\begin{abstract}
\textbf{
This study aims to optimise the ‘spray and wait’ protocol in delay tolerant networks (DTNs) to improve the performance of information transmission in emergency situations, especially in car accident scenarios. Due to the intermittent connectivity and dynamic environment of DTNs, traditional routing protocols often do not work effectively. In this study, a machine learning method called random forest was used to identify ‘high-quality’ nodes. High-quality’ nodes refer to those with high message delivery success rates and optimal paths. The high-quality node data was filtered according to the node report of successful transmission generated by the One simulator. The node contact report generated by another One simulator was used to calculate the data of the three feature vectors required for training the model. The feature vectors and the high-quality node data were then fed into the model to train the random forest model, which was then able to identify high-quality nodes. The simulation experiment was carried out in the ONE simulator in the Helsinki city centre, with two categories of weekday and holiday scenarios, each with a different number of nodes. Three groups were set up in each category: the original unmodified group, the group with high-quality nodes, and the group with random nodes. The results show that this method of loading high-quality nodes significantly improves the performance of the protocol, increasing the success rate of information transmission and reducing latency. This study not only confirms the feasibility of using advanced machine learning techniques to improve DTN routing protocols, but also lays the foundation for future innovations in emergency communication network management.
}
\end{abstract}

\section{ Introduction }
In modern communication networks, in the face of network disruption, congestion, or disaster, the traditional connection-based network model often fails to ensure the effective transmission of information. Fault-tolerant networks(DTNs)are emerging as an important solution for complex communication scenarios due to their excellent performance in the face of unstable or intermittent network connections. DTN networks \cite{7060480} employ a store-and-forward mechanism, whereby information is persistently stored in the intermediate nodes of the DTN. In this way, it can transmit information efficiently even in the face of intermittent network connections. They are widely used in emergency scenarios such as natural disasters and car accidents. As part of the Smart City project in the city of Porto, Portugal, a number of DTN protocols have been used in urban bus networks to address the problem of high mobility and low connection density between vehicles. Sensor data is transmitted through these networks to a central server for city management and environmental monitoring \cite{monteiro2015lessons}.

The focus of this research is to improve the performance of DTN networks in emergency situations, especially in the specific scenario of car accidents. The data involved is geo-temporal, dynamic and complex. These data characteristics make information delivery variable and unpredictable. In such situations, timely and reliable transmission of information is important for coordinating rescue operations and providing early warning to surrounding pedestrians and vehicles. In this scenario spray-and-wait protocol is a useful routing strategy for information delivery in DTN networks because of its simplicity and effectiveness in reducing transmission delays. The protocol ensures that information is delivered to the target node as soon as possible by propagating copies of the information among multiple nodes. However, in high-risk and unstable environments such as car accident scenes, the basic implementation of the Spray and Wait protocol may not be successful in delivering information in a timely manner.Tornell et al. (2015) published an application of Delay Tolerant Networks (DTNs) in in-vehicle networks, which demonstrated that by optimising the DTN routing strategy ( such as the Spray and Wait protocol), the success rate of message delivery in high-risk and unstable environments can be significantly improved \cite{tornell2014dtn}. The core idea of my research is to identify ‘high quality’ nodes, i.e., those nodes that have a higher probability of successfully delivering a message and a shorter path to successfully deliver a message, based on specific characteristics observed in the simulations. These nodes will be predicted by the Random Forest model. By increasing the number of message copies carried by these high quality nodes, the overall message delivery success rate of a DTN network can be significantly improved, especially in dynamic and unpredictable environments such as car accident scenes.
The rest of the paper is organised as follows: section 2 discusses global research trends in improving the performance of routing protocols in fault-tolerant networks (DTNs). Section 3 introduces the basic concepts of DTN networks and optimisation of the protocol. Section 4 describes the experimental design for setting up an emergency scenario (e.g., a car crash in Helsinki). Section 5 discusses the experimental results and analyses the reasons. Section 6 summarises the main results of the study and future research directions.

\section{Literature review}
Improving the performance of routing protocols in the field of Delay Tolerant Networks (DTNs) has been the focus of research worldwide. Many studies have introduced machine learning algorithms and other advanced techniques to address the challenge of frequent outages in these networks. In this paper, I review some important research in this area in recent years to lay the foundation for my experiments.

Firstly, the application of machine learning in DTN routing optimisation involves the use of machine learning techniques to improve the effectiveness of the protocols, such as higher data transmission efficiency and lower latency. Dudukovich, Hylton and Papachristou (2017) explored the use of a hybrid machine learning approach using Reinforcement Learning and Bayesian Learning in order to augment the performance of DTNs such as Spatial Networking routing decisions in networks to make protocols more adaptable to network changes \cite{8124902}. This study showed the potential of machine learning for adaptation but did not go into depth on how to optimise the performance of DTN networks.George and Santhosh (2021) subsequently used machine learning classifiers to improve routing decisions \cite{9640863}. Their research explored various machine learning algorithms such as decision trees and naïve Bayes to predict the best routing paths in DTN networks to optimise the success and efficiency of data transmission. This approach improves the performance of the system but does not adequately consider the energy efficiency management of the network.Radenkovic and Huynh (2020) further extended the application of machine learning in DTNs \cite{radenkovic2020cognitive}. The introduction of a deep reinforcement learning approach to optimise content caching at the network edge demonstrated the utility and effectiveness of AI techniques in improving the performance of network protocols. Through intelligent decision making, this approach improves data access speed and reduces network latency, especially during peak traffic periods.Another paper by Radenkovic and Huynh (2020) investigated how managing energy allocation using reinforcement learning can be effective in improving network energy efficiency \cite{radenkovic2020ene}. By dynamically adjusting the energy allocation according to the network conditions, the strategy optimises traffic management and network reliability and demonstrates that accurate energy allocation management can significantly improve network performance. This suggests that while existing research has optimised data transmission and energy efficiency through a variety of intelligent approaches, maintaining the stability of these measures in unstable network environments remains a challenge.

In addition, research has focused on various techniques to improve the performance of the Spray and Wait routing protocol to improve its performance in DTN environments.Kim et al. (2010) proposed a composite approach to optimise the Spray and Wait protocol by acknowledging messages in the wait phase and adjusting the message forwarding policy to address the inefficiency of the protocol in processing messages in the waiting phase \cite{5665178}.Cui et al. (2022) proposed an improved Spray and Wait routing algorithm based on social relationships among nodes \cite{9973667}. By analysing the social circles of nodes to dynamically allocate message copies, they improved the efficiency and success rate of message delivery. This shows that the performance of protocol routing can be improved by studying nodes. However, the dynamics and complexity of social relationships may affect the long-term stability of this approach. Not only in simulation, the application of machine learning in DTN routing optimisation has also been validated in real-world emergency scenarios. For example, Radenkovic and Huynh (2020) proposed a cross-layer emergency communication framework that is particularly applicable to forest fire scenes near cities. Their study showed that communication at forest fire scenes is characterised by high latency and low reliability. Thus, this illustrates the importance of optimising routing protocols to address these challenges \cite{huynh2017novel}. 

In conclusion, by studying and improving existing DTN routing protocols and performing detailed performance evaluation using advanced simulation platforms such as the One simulator, these study not only deepens the understanding of DTN communication in emergency scenarios, but also provides a reference for the design and optimisation of future network protocols. Especially in emergency situations, such as at the scene of a car accident, it is crucial to transmit information quickly and efficiently. However, these studies still have some shortcomings. In this paper, DTN routing protocols will be investigated and improved to ensure the successful and effective transmission of information in complex and unstable network environments. We introduce a random forest model to identify high-quality nodes for optimising the Spray and wait protocol. Previous research has made limited use of machine learning models to improve protocols. Even when improvements have been made, they have been to relatively simple protocols such as Epidemic. Previous research has mostly aimed to optimise energy consumption, whereas we aim to improve the success rate of message transmission and reduce latency.

\section{Delay Tolerant Networks background exploration and model optimization}

\subsection{overview of delay tolerant networks}
Delay Tolerant Networks (DTNs) are networks that incur significant and unpredictable delays, allowing routing in networks with unstable end-to-end paths. Examples include network environments with high node mobility, low node density, short radio distance, and high environmental interference and obstacles \cite{agoujil2016stochastic}. These environments may exist in backward areas or where many basic network facilities have been destroyed by natural disasters or military strikes \cite{davies2009dtn}.DTNs can also be based on mobile nodes such as vehicles, pedestrians, or underwater robots and can be applied in a variety of environments.

The first is a forwarding-based protocol, which is one of the simplest forwarding schemes in which a single node holds each message. At any given time, there is only one copy of the message in the network, so when forwarding a message, the receiving node also needs to have the ability to store the message. Next is replication based protocols, which propagate information throughout the network by replicating it. When a node encounters another node during the transmission of information, it forwards the information while keeping its own copy of the information. This is different from the forwarding based protocol where the forwarding node deletes its own local copy of the message.

One technology that should not be overlooked in such networks is CafRep technology, which represents the latest development in code congestion and data transfer in smart edge networks. CafRep technology first proposes a fully decentralised publish/subscribe model. This means that each node can independently publish information or subscribe to topics of interest. This model enhances the robustness of the network and effectively solves the difficulties of information transmission in opportunistic and delay-tolerant networks, as well as the problem of unstable connections. Also, this technology leverages the concept of edge computing, allowing data to be processed and decisions to be made close to where the data is generated. This not only reduces latency, but also increases the efficiency and reliability of data transmission. Due to its decentralised nature, CafRep is well suited for use in DTN and opportunistic network environments, where it can ensure efficient data dissemination in loaded and dynamic network environments by managing network congestion and optimising data paths \cite{radenkovic2012efficient}.

\subsection{One simulator}
In this study, the ONE simulator has been chosen as the key tool to simulate node behaviour and communication patterns in a DTN (Delay Tolerant Network) environment. The ONE simulator, full name Opportunity Network Environment, is a Java-based simulation tool specifically designed for evaluating DTN routing and application protocols. At its core, it is an agent-based discrete event simulation engine that supports highly customisable network parameter settings such as number of nodes, mobility modes, communication ranges, etc., making it ideally suited for simulating emergency communication scenarios such as car accident scenes\cite{sharma2016configuration}. Its structure is shown in the Fig 1 below\cite{keranen2009one}.

\begin{figure}[h]
    \centering
    \includegraphics[width=1.0\linewidth]{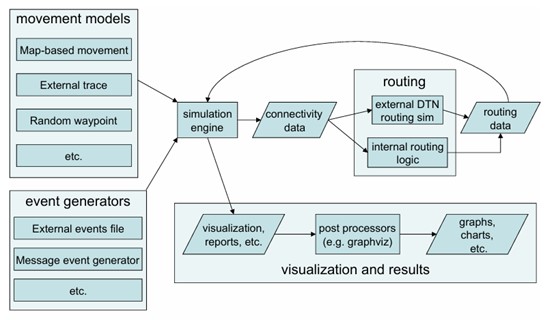}
    \caption{Overview of One simulator}
    \label{fig:1}
\end{figure}

The simulator has been widely used in academic research and has proven to be a good support for highly dynamic network environments. The powerful features of ONE include node movement, interaction between nodes, routing and message processing, and complex modelling capabilities. It also provides rich data logging capabilities that accurately capture and record every node interaction and message event in the network. The output obtained from ONE simulation can be analysed graphically using graphing tools such as Graphviz \cite{garg2021performance}. This is essential for data collection and subsequent performance analysis in this study.
In addition, the ONE simulator\cite{liu2013report} is equipped with efficient visualisation, reporting and post-processing tools which are used to collect and analyse the experimental results, enhancing the intuitiveness and interpretability of the study. Users can also present experimental results through an interactive Graphical User Interface (GUI) or by generating graphs of the data collected during the simulation, which facilitates the evaluation of different protocol configurations and behaviours. With this detailed and comprehensive simulation environment, the ONE simulator not only facilitates this research to efficiently simulate and analyse various communication strategies in DTN networks, but also provides an experimental basis for understanding and optimising the performance of the Spray and Wait protocol in real-world applications.

\subsection{Data extraction and pre-processing of information transmission nodes}
The experimental data sources for this training model are mainly from the two reports of One Simulator, namely ConnectivityDtnsim2Report and DeliveredMessagesReport. These reports contain geo-temporal data that reflects the spatial dynamics and complexity of node interactions in a real urban environment.
 
\begin{figure}[h]
    \centering
    \includegraphics[width=1.0\linewidth]{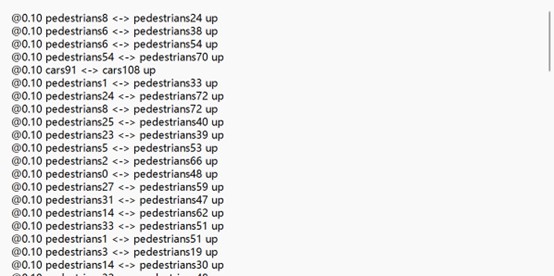}
    \caption{ConnectivityDtnsim Data}
    \label{fig:2}
\end{figure}

\begin{figure}[h]
    \centering
    \includegraphics[width=1.0\linewidth]{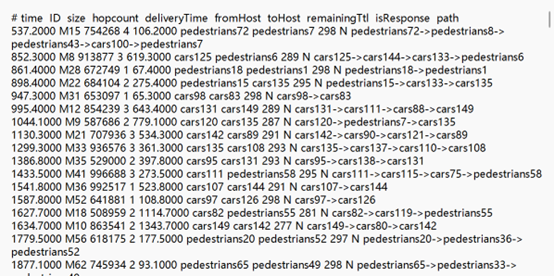}
    \caption{Delivered Messages Data}
    \label{fig:3}
\end{figure}

The data in Fig 2 can be used as training features for the random forest model after filtering. The simulator report contains data on the time nodes are connected and disconnected. Ten reports are generated, representing different situations over ten days. The number of nodes varies from day to day, which prevents the trained model from having poor generalisation ability.

The data in Fig 3 is used as high-quality nodes for the random forest model after filtering, i.e. the nodes are labelled as high quality. This report is also generated in ten copies, representing different situations in ten days, corresponding to the training features. The data contained in this simulator report mainly includes the main information of information transmission, including the time, size, path, and message survival time of information transmission. This shows that the nodes in the fromHost column have successfully transmitted information.

\subsection{Construction and application of random forest models}
The Random Forest method represents an integrated learning approach that enhances overall prediction accuracy and model robustness. This is achieved by constructing multiple decision trees and subsequently merging their predictions. This method allows each tree to grow independently on samples drawn from the original dataset via a self-sampling method, thus maintaining sample diversity. At each node of the decision tree in the Random Forest model, a random subset of features is selected from the full set of available features in order to make a split decision. This random feature selection mechanism not only increases the model's adaptability to the data, but also effectively reduces the risk of overfitting.
When training a random forest, \cite{biau2016random} each tree is allowed to grow as completely as possible without pruning, and while this makes it possible for the model to perform overfitting on the training data, the effect of this overfitting is cancelled out overall by integrating the prediction results from multiple trees. For prediction, each tree gives its prediction independently and the final classification result is determined by the output of the majority tree. In case of regression tasks, it is determined by the average of all tree outputs. The random forest model construction process is shown in Fig 4.

\begin{figure}[h]
    \centering
    \includegraphics[width=1.0\linewidth]{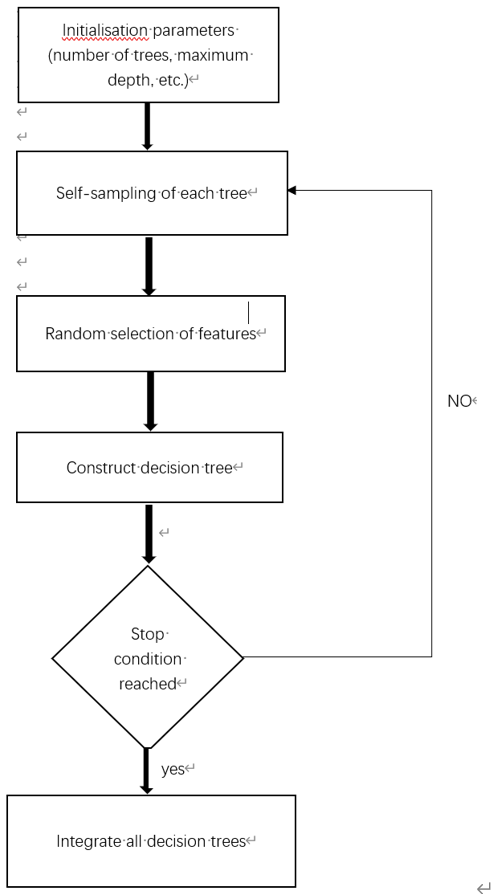}
    \caption{Random forest model construction process}
    \label{fig:4}
\end{figure}

In this study, the random forest model is used to identify high quality nodes in the DTN network. The model predicts which nodes are high quality nodes by analysing features such as the frequency of contact between nodes, the number of contacts between different nodes and the duration of connections between nodes. These features are extracted from detailed behavioural observations of the network and can effectively reflect the transmission efficiency and connection stability of nodes in the network. Using these data, the random forest model not only identifies high-quality nodes, but also provides a scientific basis for the optimisation of the Spray and Wait protocol. By increasing the number of message copies of these high-quality nodes, the study successfully enhances the success rate of message delivery in emergency situations, such as a car accident scene, thus optimising the overall performance of the network.

The model utilises data filtered by the DeliveredMessagesReport as high-quality node labels. The data filtered based on ConnectivityDtnsim2Report is used for feature input. In the training phase of the model, ten corresponding sets of data were selected for training. The ten sets of data represent the high quality node data from the first day to the tenth day and the data for feature input from the first day to the tenth day, which correspond to the ten days. 70\% of all the data is divided into training set and 30\% is divided into testing set. In order to ensure the reliability and validity of the random forest model in predicting high quality nodes, this experiment will evaluate the model by the following four metrics show in table \textnormal{I} below.

\begin{table}[h]
    \centering
    \caption{Random forest evaluation index}
    \begin{tabular}{|p{3cm}|p{4cm}|}
        \hline
        \textbf{Metrics} & \textbf{Explanation} \\
        \hline
        Accuracy & This is the most direct performance indicator, which shows the ratio of the number of correctly predicted node states (high quality or non-high quality) to the total number of predictions. The higher the accuracy, the better the overall prediction of the model. \\
        \hline
        Precision & Precision is the proportion of high-quality nodes that the model predicts to be high-quality. It measures the accuracy of the model in predicting the positive class (high-quality nodes). \\
        \hline
        Recall & The recall rate is the proportion of the number of nodes correctly identified as high-quality nodes in the model to the total number of actual high-quality nodes. This indicator is particularly important for emergency communication scenarios, because missing key nodes can lead to inefficient communication. \\
        \hline
        F1 Score & The F1 score represents a harmonic mean of precision and recall. It provides a means of balancing the two, which is particularly suitable for cases of class imbalance. \\
        \hline
    \end{tabular}
    \label{tab:random_forest_evaluation}
\end{table}

\subsection{The operating principle of the Spray and Wait protocol and its optimization}

Spray and Wait protocol is a protocol in DTN that ‘sprays’ multiple copies of a message and waits for one of the nodes in the network to deliver the message to its destination \cite{9640863}. This protocol increases the success rate of message delivery through multiple copies and reduces the wastage of network resources. The main variants of the protocol are Vanilla and binary versions \cite{das2016fibonary}.
The Spray and Wait protocol was chosen for optimisation because of its simple structure, low resource consumption and adaptability to dynamic and unstable network environments. In addition, the protocol does not rely on continuous network connectivity, is suitable for DTN environments, and is easy to test in various simulations and real devices. The protocol is particularly suitable for rapid information dissemination in emergency situations, such as car accidents, where its broadcasting characteristics can quickly cover a large area and ensure that people around the area receive emergency information in a timely manner.

\subsection{Practical application of the random forest model in protocol optimization}
The Random Forest model was chosen to optimise the ‘Spray and Wait’ protocol because of its ability to handle multidimensional data, strong generalisation and efficient performance \cite{tekouabou2022towards}. Wang and Radenkovic (2024) stated that the Random Forest model has strong predictive power in response to network changes \cite{wang2024optimization}. Therefore, the model solves the resource allocation challenges faced by the Spray and Wait protocol by predicting the optimal node characteristics to improve the transmission success rate of the DTN protocol.

The model is able to analyse the characteristics of nodes such as connection stability, transmission success and mobility and captures the relationship between these complex characteristics and the success rate of message delivery. In the protocol, Random Forest can optimise the allocation of message copies by accurately predicting node efficiency and allocating more copies to nodes with high transmission success rates, thus improving overall network performance and reducing resource wastage.

The implementation of the protocol optimisation consists of modifying the One simulator's ‘Spray and Wait’ protocol code so that high quality nodes carry double the number of message copies. Protocol performance improvements were evaluated by processing the ConnectivityDtnsim2Report data generated by the simulator and predicting high-quality nodes using the Random Forest model, which were then entered into the configuration file and subjected to a secondary simulation without changing other conditions.

Ultimately, by analysing the MessageStatsReport generated by the One simulator, performance metrics (e.g., delivery probability, latency, etc.) can be used to evaluate the optimisation of the Spray and Wait protocol in real communication scenarios and it shows in table \textnormal{II} below.

\begin{table}[h]
    \centering
    \caption{Protocol performance evaluation}
    \begin{tabular}{|c|p{5cm}|}
        \hline
        \textbf{Metrics} & \textbf{Explanation} \\
        \hline
        Delivery prob & Measures the proportion of all generated messages that successfully reach their destination. \\
        \hline
        Overhead ratio & Indicates the ratio of the total number of transmission attempts made by the network to the number of successfully delivered messages in order to achieve a successful message delivery. \\
        \hline
        Latency avg & Indicates the average time required for a message to successfully reach its destination after it is created and sent out. This indicator generally reflects the speed and efficiency of information delivery in the network. \\
        \hline
        Buffertime avg & Indicates the time a message spends in the cache of a network node. This indicator reflects the efficiency of the network in processing messages. \\
        \hline
    \end{tabular}
    \label{tab:protocol_performance_evaluation}
\end{table}

\section{Experiment design and implementation}
\subsection{emergency smart city scenario design}

In this article, we set up an emergency scenario that allows the use of DTN networks and improves the performance of the ‘spray and wait’ protocol in this emergency scenario by increasing the number of copies carried by high-quality nodes.
The emergency scenario set up in this article is a car accident scenario in the city of Helsinki. The scenario has diverse and easily changing characteristics, making the effective transmission of information particularly difficult. The variable is set to the number of nodes, and the number of nodes on holidays will be higher than usual. The car accident scene continuously sends information as a fixed node. The surrounding pedestrians and vehicles receive the information from the crash site as mobile nodes and pass it on to other pedestrians and vehicles in the city to warn them. The information is also passed on to the emergency services so that they can respond to possible casualties. The mobile nodes make the network topology constantly change due to their uncertain movement paths, which may result in connections lasting only a short time, making information transmission more difficult. In addition, information needs to be transmitted quickly and accurately from the crash site to nearby pedestrians and vehicles. Due to the high mobility of the nodes, the transmission path is often interrupted and delayed. This dynamic environment significantly reduces the reliability of information transmission. As described in \cite{radenkovic2019enabling}, there are significant challenges due to physical obstructions in an urban environment, especially in densely populated city centres. This results in reduced communication performance and reliability between connected vehicles and drones. So in realistic scenarios, urban structures such as tall buildings can also make the transmission of useful information more difficult due to factors such as signal blockage and network interference.

\subsection{Experimental process for improving the performance of the Spray and Wait protocol using a random forest model }

\begin{figure}[h]
    \centering
    \includegraphics[width=1.0\linewidth]{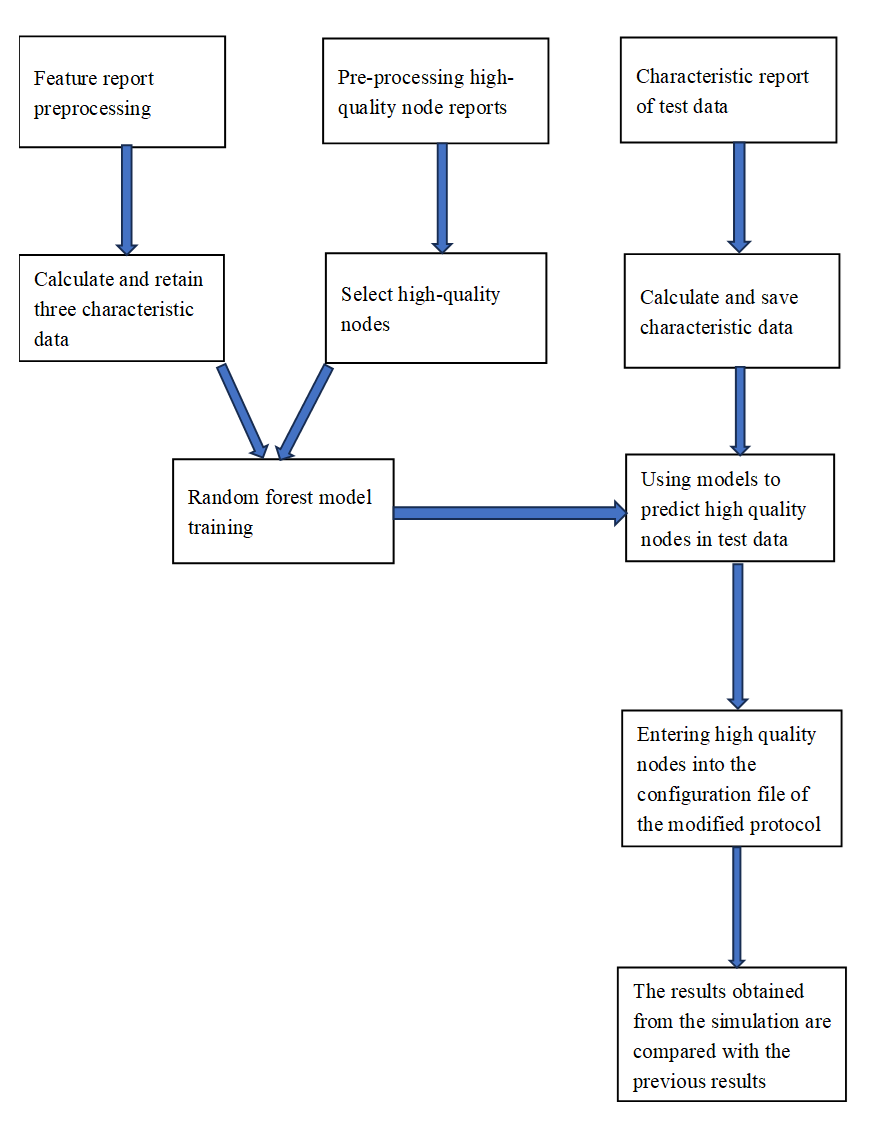}
    \caption{experimental flow chart}
    \label{fig:5}
\end{figure}

\subsection{Experiment set up}
\subsubsection{Scene Node Setting}
In this experiment, four groups of nodes were defined for simulation.
\begin{table}[h]
    \centering
    \caption{Node information}
    \begin{tabular}{|c|p{5cm}|}
        \hline
        \textbf{Group number} & \textbf{Group ID} \\
        \hline
        Group1 & Pedestrians \\
        \hline
        Group2 & Cars \\
        \hline
        Group3 & Accidents \\
        \hline
        Group4 & Rescue centres \\
        \hline
    \end{tabular}
    \label{tab:group}
\end{table}

First of all, there are some basic settings for the nodes groups shown like Fig 6 and the explanation of the settingsis shown like Table \textnormal{IV}.
 
\begin{figure}[h]
    \centering
    \includegraphics[width=1.0\linewidth]{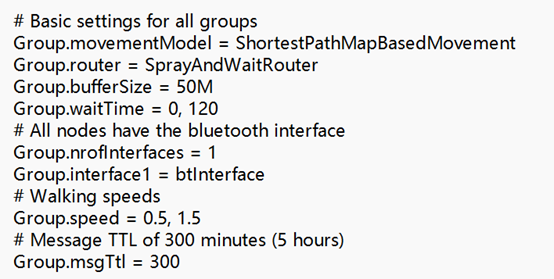}
    \caption{basic settings of groups}
    \label{fig:6}
\end{figure}

\begin{table}[h]
    \centering
    \caption{details of basic settings}
    \begin{tabular}{|c|p{5cm}|}
        
        \hline
        basic settings for all groups \\
        \hline
        Group.movementModel=ShortestPathMapBasedMovement \\
        \hline
        Group.router = Spray and WaitRouter \\
        \hline
        Group.bufferSize = 50M \\
        \hline
        Group.waitTime = 0, 120 \\
        \hline
        Group.nrofInterfaces = 1 \\
        \hline
        Group.interface1 = btInterface \\
        \hline
        Group.speed = 0.5, 1.5 \\
        \hline 
        Group.msgTtl = 300 \\
        \hline  
    \end{tabular}
    \label{tab:basic settings}
\end{table}
There are also some parameter settings that are fixed in this scenario shown in table \textnormal{V} below.
\begin{table}[h]
    \centering
    \caption{Explanation of Fix Parameters}
    \begin{tabular}{|c|p{3cm}|}
        \hline
        \textbf{Parameter} & \textbf{Value} \\
        \hline
        Scenario.endTime & 43200 \\
        \hline
        btInterface.transmitSpeed & 250k \\
        \hline
        btInterface.transmitRange & 30 \\
        \hline
        Car speed & 2.7, 13.9 \\
        \hline
        Number of accident & 1 \\
        \hline
        Number of rescue centers & 2 \\
        \hline
        Message sizes & 500kB- 1MB \\
        \hline
       Message creation interval in seconds & 25-35 \\
        \hline 
         
    \end{tabular}
    \label{tab:fix}
\end{table}

\subsubsection{Random forest model training}
Firstly, ten different node combinations were used to simulate the current scenario using the unmodified Spray and Wait protocol ten times to obtain ten different copies of the ConnectivityDtnsim2Report and DeliveredMessagesReport. In the ten groups of nodes, the number of nodes for car accidents is fixed at 1, the number of nodes for rescue centres is fixed at 2, and the number of nodes for pedestrians and cars varies for each group. The table \textnormal{VI} below shows the number of pedestrian as well as car nodes for each group.
\begin{table}[h]
    \centering
    \caption{ Node combination of different groups}
    \begin{tabular}{|c|p{2cm}|}
        \hline
        \textbf{pedestrians} & \textbf{cars} \\
        \hline
        50 & 60 \\
        \hline
        50 & 65 \\
        \hline
        50 & 70 \\
        \hline
        55 & 65 \\
        \hline
        55 & 60 \\
        \hline
        60 & 70 \\
        \hline
        60 & 80 \\
        \hline
        60 & 60 \\
        \hline 
        65 & 75 \\
        \hline
        70 & 80 \\
        \hline
         
    \end{tabular}
    \label{tab:Node combination}
\end{table}

Extracting useful information from the reports generated by the simulator to get high quality nodes. High-quality nodes are those that are more likely to successfully transmit information over a shorter path. We chose the following method to select high-quality nodes: First, we filtered out source nodes with shorter transmission path lengths. Then, we filtered out the nodes with a high number of successful transmissions (more than 8 successful transmissions, up to 16 successful transmissions for a node). We randomly select 50\% of the nodes from the list and mark them as high quality nodes. Some of the high quality nodes are shown in Fig 7 below. It shows that the transmission paths of these high-quality nodes are very short and the number of successful passes is high.
 
\begin{figure}[h]
    \centering
    \includegraphics[width=0.7\linewidth]{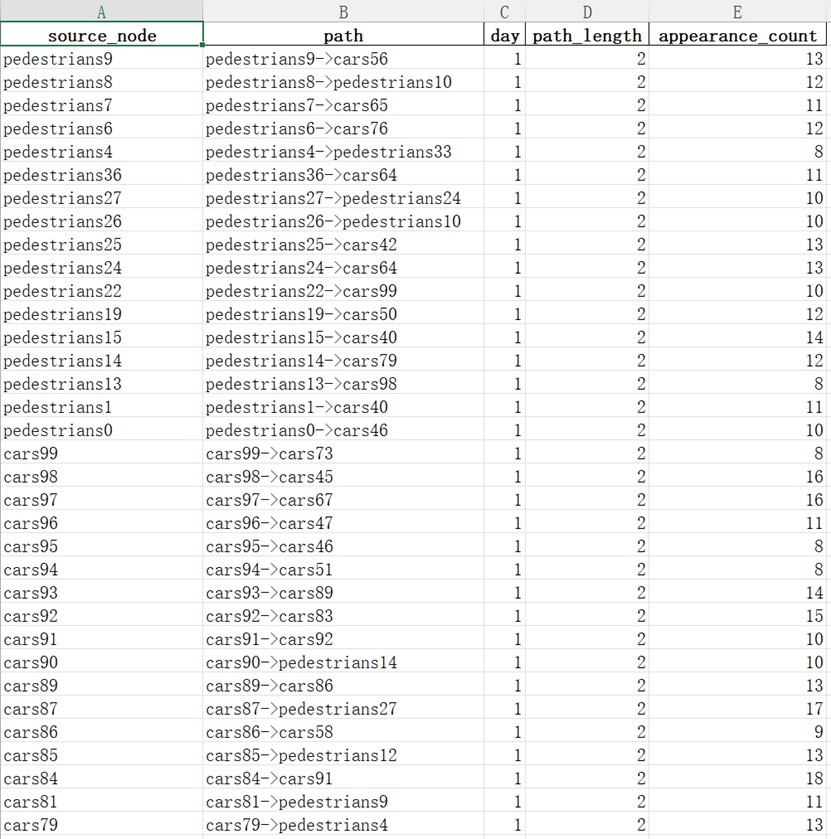}
    \caption{list of high-quality nodes}
    \label{fig:6}
\end{figure}
The three features required to train the model are then extracted from the report generated by the simulator, and the extracted information is shown in Fig 8 contact\_frequency represents the number of times the source node has been in communicating with other nodes, degree represents how many different nodes the source node has been in contact with, and duration represents the average duration of the connection between the source node and other nodes. These features are very important for complex and changeable data.    
\begin{figure}[h]
    \centering
    \includegraphics[width=0.5\linewidth]{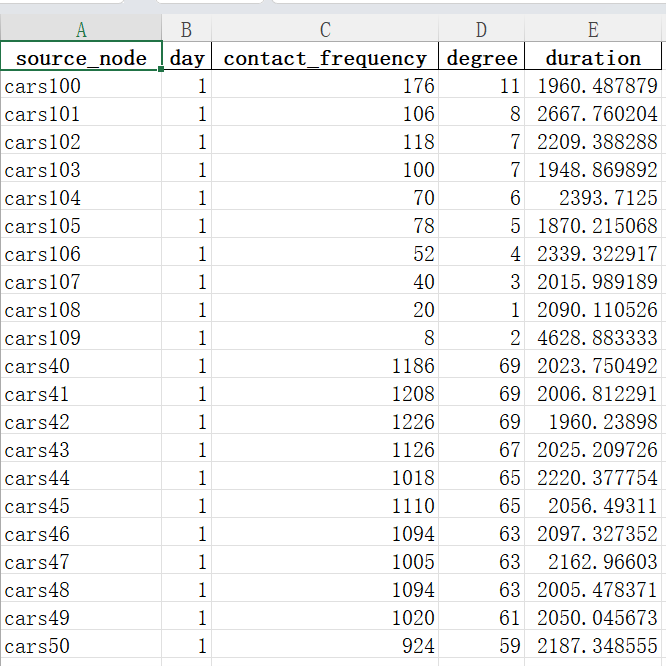}
    \caption{feature extration indormation list}
    \label{fig:7}
\end{figure}

Load feature data and high quality node data. The high quality nodes are then labelled. Select the features and labels then train the random forest model. The model training setup is shown below. 70\% of the data is used as training set and 30\% as test set. The number of trees is set to 200 and the maximum depth.

The performance of the trained model is shown in the table \textnormal{VII} below, with 0 representing non-high quality nodes and 1 representing high quality nodes.

\begin{table}[h]
    \centering
    \caption{Classification Report}
    \begin{tabular}{|c|c|c|c|p{2cm}|}
        \hline
        & \textbf{Precision} & \textbf{Recall} & \textbf{F1-score} & \textbf{Support} \\
        \hline
        0.0 & 0.75 & 0.54 & 0.63 & 218 \\
        \hline
        1.0 & 0.72 & 0.88 & 0.79 & 304 \\
        \hline
    \end{tabular}
    \label{tab:classification_report}
\end{table}

\subsection{Spray and Wait Protocol Enhancement}

Firstly, modify the code of the Spray and Wait protocol so that the protocol can get the values related to high\_quality\_nodes in the configuration file, and then store the node names in the highQualityNodeNames variable, so that it can subsequently judge whether a node belongs to a high quality node through this collection.
Add a judgement module to the protocol. If node belongs to high quality node then number of replicas carried by the node × 2.
Because the getName() method is not defined in the DTNHost.java file, it needs to be added in that file.

\subsection{Validation protocols enhance effectiveness}
In order to verify the change in performance after the protocol enhancement, two sets of control experiments were set up, before and after changing the protocol, during weekdays and holidays, respectively, and the change in simulation results. The difference between weekdays and holidays is only the difference in the number of pedestrian nodes and car nodes.They show in table \textnormal{VIII} below.

\begin{table}[h]
    \centering
    \caption{ Workday and holiday nodes combination}
    \begin{tabular}{|c|c|p{2cm}|}
       
        \hline
        &weekdays & holidays \\
        \hline
        pedestrians & 60 & 75 \\
        \hline
        Cars & 70 & 85  \\
        \hline

    \end{tabular}
    \label{tab:workdays and holidays}
\end{table}

Firstly, the working day group is observed first and after setting the number of nodes for pedestrians and cars in the configuration file, the simulation is performed using the unmodified Spray and Wait protocol, which is saved as simulation result 1. In the result, the feature data of the nodes are extracted, and the feature data are fed into the trained model, through which the high quality nodes are predicted.Some of the selected data shown in the Fig 9 below. These high quality nodes are entered into the highQualityNode item in the profile of the modified Spray and Wait protocol, and the simulation is re-run with all other conditions unchanged and saved as simulation result 2. Finally, in the protocol configuration file, enter the same number of nodes as in result 2 in the highQualityNode item, and then simulate with all other conditions remaining unchanged, and save the simulation as result 3. A comparison of the three results is shown in the table\textnormal{IX} and table \textnormal{X}  below.

\begin{figure}[h]
    \centering
    \includegraphics[width=0.7\linewidth]{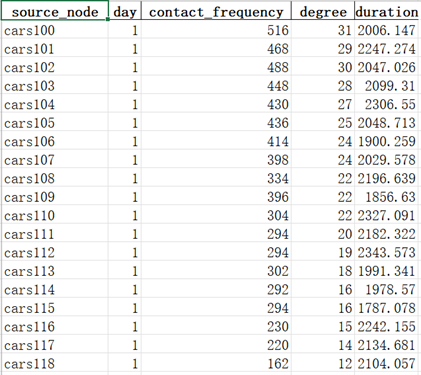}
    \caption{Selected data extracted}
    \label{fig:12}
\end{figure}

\begin{table}[h]
    \centering
    \caption{ Workday Group Protocol Simulation Results}
    \begin{tabular}{|c|c|c|c|p{2cm}|}
       
        \hline
        weekdays & Original(1) & High-quality nodes(2) & Random node(3)  \\
        \hline
        Delivery prob & 0.7262 & 0.7861 & 0.7684 \\
        \hline
        Overhead ratio & 6.7627 & 9.2981 & 9.4681  \\
        \hline
        Latency avg & 3016.9181 & 2821.7711 & 2815.8483 \\
        \hline
        Buffertime avg & 17779.3085 & 17582.3337 & 17660.5192 \\
        \hline

    \end{tabular}
    \label{tab:workdays result}
\end{table}

\begin{table}[h]
    \centering
    \caption{ Holiday Group Protocol Simulation Results}
    \begin{tabular}{|c|c|c|c|p{2cm}|}
       
        \hline
        holidays & Original(1) & High-quality nodes(2) & Random node(3) \\
        \hline
        Delivery prob & 0.7207 & 0.7759 & 0.7589 \\
        \hline
        Overhead ratio & 6.8507 & 9.7392 & 9.8465  \\
        \hline
        Latency avg & 2977.2023 & 2613.7557 & 2636.4879 \\
        \hline
        Buffertime avg & 17847.0380 & 17767.6789 & 17780.8958 \\
        \hline

    \end{tabular}
    \label{tab:holiday group}
\end{table}

To evaluate performance, I will use a rich multi-criteria approach, focusing on four key evaluation metrics: delivery probability, overhead ratio, average latency, and average buffering time. Below, I will analyze each metric in turn. Fig 10 compares the probability of message delivery during weekdays and holidays for the original protocol, the protocol using high-quality nodes, and the protocol using random nodes. Fig 11 shows their overhead ratio. Fig 12 compares their Latency avg. Fig 13 shows their Buffertime avg.

 \begin{figure}[h!]
    \centering
    \includegraphics[width=0.5\linewidth]{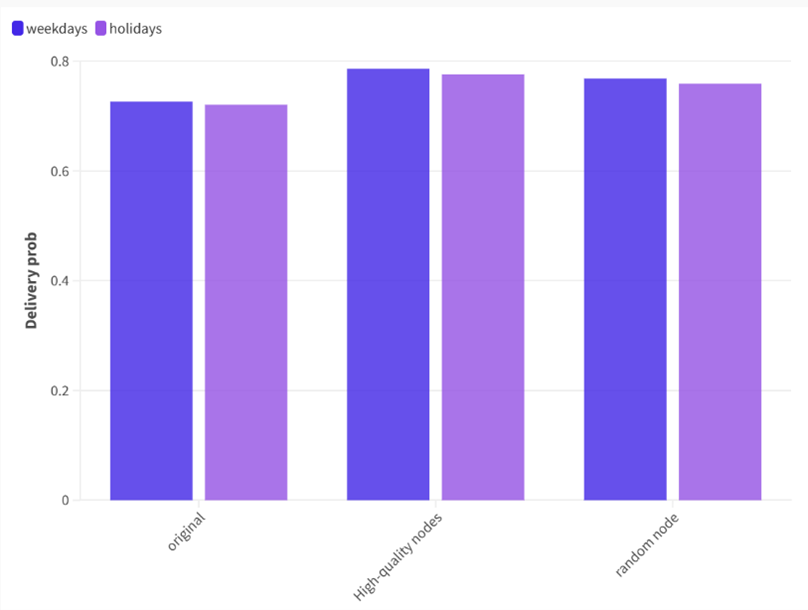}
    \caption{Delivery probability result comparison}
    \label{fig:13}
\end{figure}

\begin{figure}[h!]
    \centering
    \includegraphics[width=0.5\linewidth]{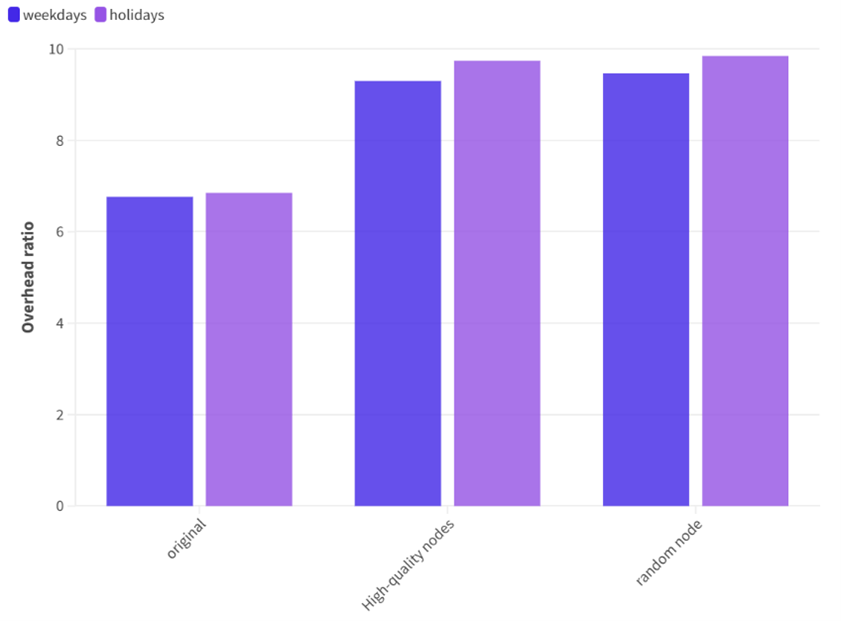}
    \caption{Overhead ratio result comparison}
    \label{fig:14}
\end{figure}

\begin{figure}[h!]
    \centering
    \includegraphics[width=0.5\linewidth]{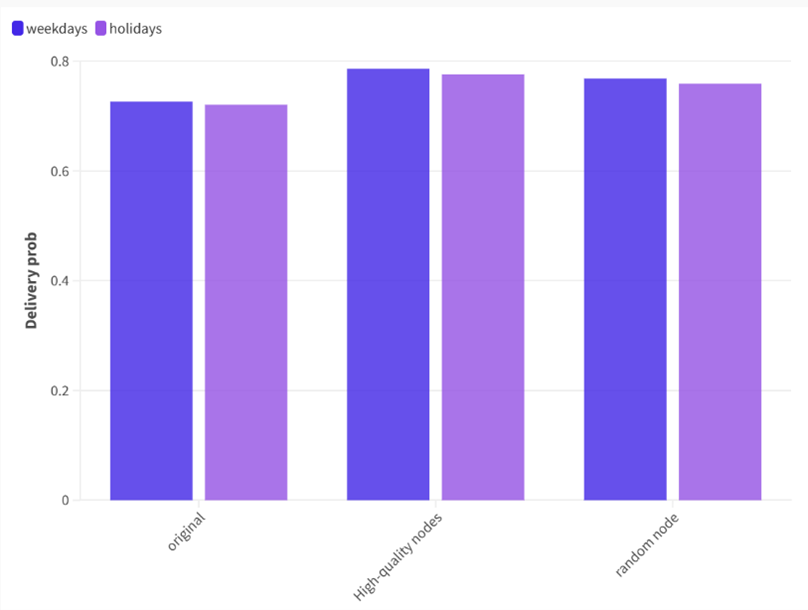}
    \caption{Latency avg result comparison
}
    \label{fig:15}
\end{figure}

\begin{figure}[h!]
    \centering
    \includegraphics[width=0.5\linewidth]{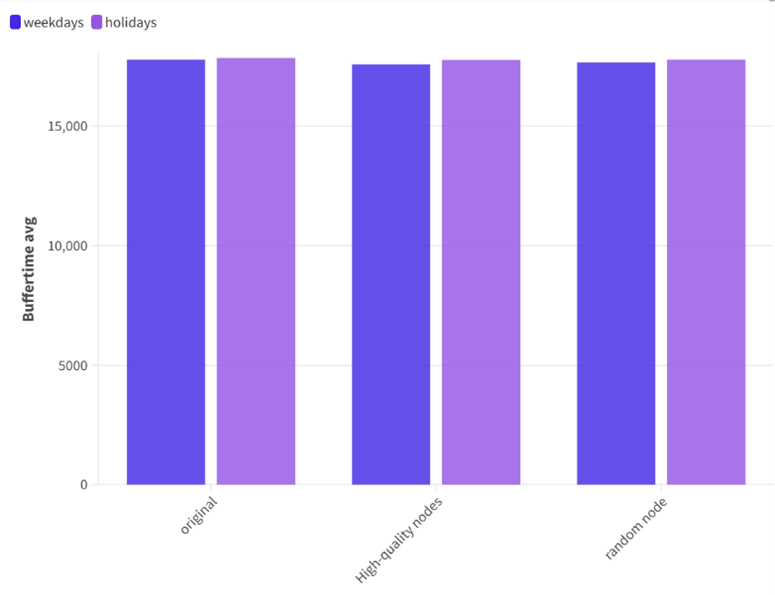}
    \caption{Buffertime avg result comparison}
    \label{fig:16}
\end{figure}

\section{Discussion}\
\subsection{Spray and Wait Protocol performance improvement}
The experimental results show that the introduction of high-quality nodes increases the success rate of message transmission by 6\% on weekdays and 5\% on holidays. This indicates that selecting nodes with high transmission efficiency through the random forest model can significantly enhance the transmission capacity of the network. The optimisation effect is more pronounced on weekdays, indicating that the optimisation effect may be affected by the dynamic network environment when there are more nodes. In addition, the average latency and caching time of message delivery are reduced, indicating that the network response speed and processing efficiency are improved. In emergency situations, these improvements ensure that critical information can be delivered in a timely manner. Despite the increased network overhead, the moderate consumption of resources is justified in emergency situations and the reduced latency compensates the overhead to some extent. Compared to random nodes, the introduction of high-quality nodes increased the success rate of message delivery by 2\%. This verifies the effectiveness of the random forest model in identifying high-quality nodes, and also shows that the success rate of message delivery in emergency situations can be improved by an efficient node selection strategy.
\subsection{Random forest model performance}
In selecting high-quality nodes, the experiment uses nodes with a transmission path of 2 and more than 8 successful transmissions, and selects 50\% of them as high-quality nodes. This method can effectively improve the protocol performance, although this screening criterion does not fully reflect the real transmission capability of the nodes. Future research can introduce more behavioural features to optimise node screening. The feature vectors of the model include the contact frequency of nodes, the number of contacting nodes and the connection duration. The random forest model captures complex feature relationships through multiple decision trees, effectively reducing the overfitting problem that may be caused by a single model. The model is highly scalable, but the training time is long and depends on the quality of the training dataset. In this study, the random forest model has an accuracy of 73\% and a recall of 88\%. Despite the small dataset, the predictive performance of the model is close to the application requirements of emergency scenarios, but further expansion of the dataset is needed to improve the generalisation ability.
\subsection{Practical application}
Although the protocol performs well in simulation, it still faces many challenges to be applied in real scenarios. In a real crash scenario, factors such as terrain and electronic devices may interfere with the wireless signals and affect the communication quality of the DTN network. In addition, the dynamics of vehicles and pedestrians are different from the fixed-interval movement in simulation, and this uncertainty may reduce the accuracy of the model. The processing power and resource limitations of the devices also have an impact on the communication efficiency, and the deployment of advanced DTN protocols requires investment in hardware and software development, and maintenance costs need to be considered.
Future research can verify the broad applicability of the protocols through more practical scenarios, while dynamically adjusting the model parameters to adapt to different emergencies situation.The implementation of this method can also benefit from the cost-effective and decentralised cloud infrastructure provided by low-cost mobile personal clouds \cite{radenkovic2016low}. By using mobile personal clouds, we can improve the flexibility of DTN-based emergency communication systems. At the same time, it can also maintain low operating costs in resource-constrained environments. In addition, data security and privacy protection are particularly important in real-world applications, and data encryption and anonymisation techniques should be explored in the future \cite{kattenborn2021review}. In terms of models, future research can introduce more behavioural features to further improve the predictive ability of models. Deep learning models such as CNN and RNN can handle spatio-temporal data and are suitable for dynamic network environments \cite{kattenborn2021review}\cite{li2015lstm}. LSTM can also be used for the prediction of network nodes to improve the accuracy of the model \cite{li2015lstm}. In addition, online learning methods can dynamically update the model based on real-time data to adapt it to environmental changes and enhance its practicality and robustness. In the future, smarter resource allocation strategies need to be explored to reduce resource consumption and improve transmission efficiency by adaptively adjusting the number of replicas and combining node capacity with network conditions.

\section{Conclusion}
This study aims to improve the efficiency of information transmission in emergency situations by optimising the Spray and Wait protocol in Delay Tolerant Networks (DTNs). We simulate a car crash scenario in a simulated environment and combine a random forest model to select high-quality nodes and increase the number of replicas they carry, in order to improve the transmission success rate of the network. The results show that selecting high-quality nodes significantly improves the success rate of information transmission and reduces transmission delays, and also performs well in complex network environments. Although this strategy increases network overhead, a moderate consumption of resources is reasonable and necessary in emergency situations.

Moreover, the experiment verified the potential of machine learning in dynamic and unstable environments, and the introduction of the model effectively enhanced the network performance. Future research can expand the dataset to further improve the model performance and explore intelligent resource allocation and data security techniques. Although this study is based on a simulation environment, we hope that these optimisation strategies can be verified in real networks and promote the application and development of DTN technology in future cities to cope with emergency situations.

\bibliographystyle{IEEEtran}

\bibliography{references.bib}

\end{document}